\documentclass{elsart}

\usepackage{graphicx}
\usepackage{epsfig}
\usepackage{amssymb}

\input epsf         

\begin{document}

\begin{frontmatter}

\title{A slow and dark atomic beam}
\author{B. K. Teo, T. Cubel, G. Raithel}
\address{FOCUS Center and Physics Department, University of
Michigan, 500 East University, Ann Arbor MI 48109, USA. }

\begin{abstract}
We demonstrate a method to produce a very slow atomic beam from a
vapour cell magneto-optical trap. Atoms are extracted from the
trap using the radiation pressure imbalance caused by a push beam.
An additional transfer beam placed near the center of the trap
transfers the atomic beam into an off-resonant state. The velocity
of the atomic beam has been varied by changing the intensity of
the push beam or the position of the transfer beam. The method can
be used to generate a continuous, magnetically guided atomic beam
in a dark state.

\end{abstract}

\end{frontmatter}

\section{Introduction}

Cold atomic beams are useful for atom interferometry and precision
atomic spectroscopy experiments. Cold beams were first derived
from oven sources and slowed by the radiation pressure from a
resonant laser beam which can be temporally chirped \cite{chirp}
or operated at fixed frequency in conjunction with Zeeman coils
\cite{zeemancoil}. These atomic beams have a large transverse
velocity spread because the atoms are cooled in only one
dimension. Colder atomic beams that are both transversely and
longitudinally cooled can be derived from 2D magneto-optical traps
(MOTs) and moving optical molasses
\cite{funnel1,2DMOT,funnel2,funnel3}. The mean beam velocity
produced by these 2D MOTs is tunable between 0.5-3~m/s. Atoms can
also be extracted from 3D MOTs using the Low Velocity Intense
Source (LVIS) geometry, in which a radiation pressure imbalance is
created by a small hole at the center of one of the
retro-reflecting mirror-waveplate assemblies \cite{LVIS}, or by an
additional laser beam \cite{pushLVIS1,pushLVIS2}. These sources
produce atomic beams with a mean velocity of about 15~m/s.
Recently, the propagation of cold atomic beams in magnetic guides
has been studied \cite{magguide1,magguide2,magguide3,magguide4}.

Evaporative cooling of transversely confined atomic beams is a
promising route towards the generation of a continuous source of
Bose Einstein Condensates (BEC) \cite{dalibard}. This approach to
a continuous coherent atom source differs from the method of
periodic replenishment of a static BEC reservoir
\cite{ketterlecw}, where suitably interlaced moving optical dipole
traps are used. In the latter method, large number and phase
fluctuation of the condensate can be expected. The phase and atom
number will be more stable if the whole system runs continuously,
without moving potentials and/or pulsed atomic beams. A BEC
continuously replenished by an evaporatively cooled, guided atomic
beam should therefore have much lower phase and atom number
variations.

Evaporative cooling of a magnetically guided atomic beam presents
 technical challenges. To achieve high enough densities and
collision rates, a very large flux of cold atoms (better than
$10^{9}$~s$^{-1}$) moving at very low velocities (less than 1~m/s)
is required. The atomic flux must be in a low-field-seeking state
and must be efficiently coupled into the guide. These challenges
can only be partially met when using the beam sources described in
\cite{LVIS,pushLVIS1,pushLVIS2}, which produce atomic beams with
longitudinal velocities of order 15~m/s. Moving optical molasses
combined with magneto-optic trapping can produce dense, guided
atomic beams with velocities in the 1~m/s range \cite{dalibard2}.
Other schemes to couple cold atoms into magnetic guides have been
studied in \cite{magguide3,magguide4}.

The simultaneous operation of a MOT producing an atomic beam and a
nearby magnetic  atomic-beam guide is a further challenge. In such
a system, the MOT will produce some stray light in its vicinity.
The MOT stray light will be scattered by the atoms propagating in
the atom guide, resulting in optical pumping of the atoms out of
the desired low-field-seeking magnetic sublevel. Stray light
intensities $< 1$~nW/cm$^2$ of near-resonant light (for Rb atoms)
will seriously attenuate the magnetically guided atomic beam. It
appears very hard to suppress the MOT stray light below that
intensity level. A promising way out of this dilemma is to
optically pump the atomic beam emitted from the MOT into a dark,
low-field-seeking state before the near-resonant stray light from
the MOT depletes it. In alkali atoms, a magnetic sublevel of the
lower of the two ground-state hyperfine levels can be used as such
a dark state.  To ensure a rapid enough transfer into that state,
a separate optical-pumping laser can be used. In this paper, we
use this idea to produce a slow beam of $^{87}$Rb atoms in the
$\vert F=1, m_{\rm F}=-1 \rangle $ ground-state sublevel, which is
dark with respect to the stray light of a nearby MOT (which is the
atomic-beam source). In addition to being in a dark state, the
atomic beam is also considerably slower than the bright atomic
beams in \cite{LVIS,pushLVIS1,pushLVIS2}. Experimental data from
pulsed measurements are shown, but the method can also generate
continuous, dark atomic beams. The beam may be magnetically guided
and, in future efforts, evaporatively cooled while the MOT is
operating.

\section{Extraction of Atoms from a Vapour Cell MOT}

We use a six-beam vapour-cell MOT and a push laser to produce a
$^{87}$Rb atomic beam. The radiation pressure imbalance due to the
push beam \cite{pushLVIS1,pushLVIS2} accelerates cold atoms out of
the center of the MOT. The MOT and pusher beams act on the upper
hyperfine level $F=2$ of $^{87}$Rb. The velocity of the resultant
atomic beam is determined by the intensity, detuning, and
polarization of the push laser. Since the Zeeman and the Doppler
shifts of the atoms are position- and/or velocity-dependent, the
acceleration also depends on position and velocity.  On average, a
$^{87}$Rb atom extracted from the MOT scatters about 10,000
photons from the push beam before it is pumped into the lower
hyperfine state $F=1$ via an accidental off-resonant $F=2
\rightarrow F'=2$ transition. The acceleration phase of the atom
can, however, be terminated early using an additional laser beam,
referred to as transfer beam, that optically pumps the atom into
the $F=1$ ground state. Because of the large energy spacing
between the $F=1$ and $F=2$ ground states, the push and the MOT
light is not resonant with the $F=1$ atoms. The dark atomic beam
is slower than that produced by a conventional LVIS. Also, {\em
all} atoms in the beam cease accelerating at the same location,
namely when they enter the transfer beam. This fact leads to a
velocity spread of the dark atomic beam that is less than the
velocity spread of a conventional LVIS atomic beam.

The repumper light required to operate the MOT is resonant with
the $F=1$ $\rightarrow$ $F'=2$ transition and must therefore be
sufficiently shielded from the $F=1$ dark atomic beam. There is
only a single MOT repumper beam, and that beam has a low intensity
and is quite small. Therefore, it is technically {\sl much} easier
to shield the repumper stray light from the dark atomic beam than
preventing the MOT stray light, which is near-resonant with the
$F=2$ $\rightarrow$ $F'=3$ transition, from reaching the atomic
beam. This (practical) difference in the ability to shield MOT
stray light from the atomic  beam vs. the ability to shield
repumper stray light has been a major motivation for the
development of the method presented in this paper.

\section{Experimental Setup}
\label{esetup}

\begin{figure} [hbt]
\centerline{\epsfig{figure=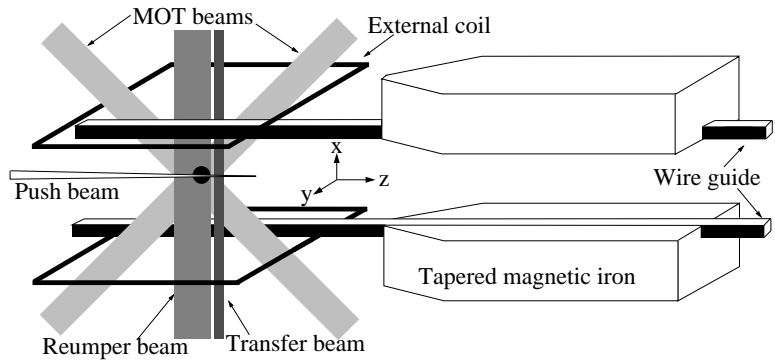,width=4.0in}}
\caption {Schematic of our magnetic guide setup. The vacuum chamber
(not shown) rests in between and parallel to the wire guide. The magnetic
iron is tapered to give a smooth change in transverse field gradient
along the guide axis.}
\label{setup}
\end{figure}

Our experimental setup has been described previously in
\cite{bouncepaper}  and is shown schematically in
Fig.~\ref{setup}. We use a two-wire magnetic atom guide. The guide
wires are separated by 2.8~cm and are placed outside a quartz
vacuum tube. The transverse magnetic field of the atom guide
varies along the guide axis ($z$ direction) due to the presence of
two tapered pieces of magnetic iron. At a current of 400~A, the
transverse magnetic field gradient increases from 70~G/cm to
320~G/cm along the guide. $^{87}$Rb atoms are captured from the
background gas in a MOT, which is formed on the axis of the guide.
The MOT laser beams are detuned by a frequency of $2\Gamma / 2 \pi
= 12$~MHz to the red of the $5S_{1/2} F=2 \rightarrow 5P_{3/2}
F'=3$ transition and have an intensity of 5~mW/cm$^2$. An
additional beam resonant with the $5S_{1/2} F=1 \rightarrow
5P_{3/2} F'=2$ repumps atoms that fall into the $F=1$ state back
into $F=2$. In our pulsed experiments, during the MOT loading
phase the guide current is reduced to 200~A, and a pair of
external magnetic coils is energized, resulting in a MOT magnetic
field with gradients -15, 26 and -11~G/cm in the $x, y$ and $z$
directions. In each loading cycle, $2\times10^7$ atoms are
collected by the MOT.

Cold atoms are extracted from the MOT using a push laser beam
propagating along the guide axis \cite{pushLVIS1,pushLVIS2}. In
the presented experiments, the guide current (see Fig.~1) is
increased from 200~A to 400~A at the time when the push laser beam
turns on, while the current in the external coils  is kept
constant. The push beam has the same detuning as the six MOT beams
and produces a radiation pressure imbalance at the MOT center that
forces the atoms out. The MOT repumper beam intersects the
$z$-axis at a right angle and is masked by a sharp edge such that
the repumper intensity abruptly drops to zero a few mm downstream
from the MOT (see Fig.~1). The transfer beam, located close to the
MOT, optically pumps the extracted atoms into the $F=1$ dark
state. The transfer beam is locked to the $5S_{1/2} F=2
\rightarrow 5P_{3/2} F'=2$ transition. The transfer beam has a
circular cross section (diameter of $\approx 2$~mm) with an
approximately constant intensity equal to the saturation intensity
($I_{\rm sat}=1.6$~mW/cm$^2$). The transfer beam interrupts the
acceleration of the atoms. While providing a well defined location
of transfer from $F=2$ to $F=1$, repeated transitions of the atoms
back and forth between these states must be avoided. Therefore,
any spatial overlap between the repumper and the transfer beam
needs to be avoided.

In our experiments we have, for convenience, used an existing
small uncoated quartz cell as our vacuum chamber. Therefore, the
amount of stray light present in the vacuum system has been
unusually high. In order to reduce the amount of repumper stray
light in the atom guiding region to an acceptable level, the MOT
repumper intensity had to be reduced to levels so low that the
number of atoms in the MOT has been significantly reduced.
Obviously, a larger vacuum chamber with dark surfaces and vacuum
windows placed outside direct line-of-sight from the guided atoms
will allow us to use a higher MOT repumper intensity.

\section{Experimental Data}

To detect the fluorescence of the atomic beam, the beam is imaged
onto a 3~mm by 3~mm photodiode (magnification approximately one).
The center of the photodiode corresponds to an object location
that has a distance of 1~cm from the MOT center. To measure the
time-of-flight distribution of the atomic beam, a mechanical
shutter is used to pulse the push beam. The onset of the push beam
extracts a pulse of atoms from the MOT. The MOT beams can be
turned off or left on continuously during the extraction. The
extracted atoms stream through the region that is imaged onto the
photodiode. Atoms that enter the detection region in the state
$F=2$ are still being accelerated by the push beam. The resultant
fluorescence is detected, and its time-dependence can be used as a
measure for the average velocity and velocity spread of the atoms.
To be able to also detect atoms propagating in the dark state
($F=1$), an additional repumping beam can be introduced into the
detection region. This repumper returns atoms propagating in the
dark state $F=1$ into the bright state $F=2$, thereby making them
detectable.

\subsection{Pulsed Measurements of Velocity}

\begin{figure} [hbt]
\vspace{0.5cm}
\centerline{\epsfig{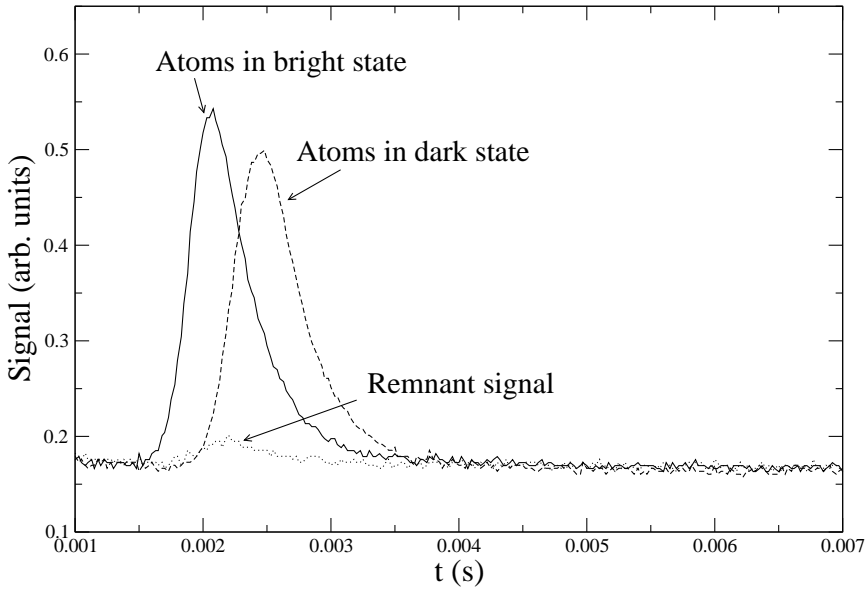}}
\vspace{0.5cm} \caption {Time-resolved measurements of the
fluorescence of a pulsed atomic beam at a distance of 1~cm from
the MOT. The push beam is turned on at $t=0$. The solid curve is
the signal without the transfer beam. When the transfer beam is
turned on, most atoms are pumped into the dark state and do not
fluoresce; the small fraction of atoms left in the bright state
produce a weak remnant signal (dotted curve). In the dashed curve,
an additional repumper beam in the detection region is used to
pump the atoms in the dark state back into the bright state. The
delay between the solid and dashed curves reflects the velocity
difference between the bright and dark atomic beams.}
\label{darkstate}
\end{figure}

The following data shows the time dependence of the fluorescence
signal for the case that the leading edge of the transfer beam is
located 3~mm from the MOT. The MOT beams are shut off before the
push beam is turned on. Without the transfer beam, most of the
atoms remain in the $F=2$ bright state and scatter light from the
push beam (solid curve in Fig.~\ref{darkstate}) while
accelerating. Assuming constant acceleration we find, based on the
average arrival time of the atoms, an average velocity of the
atoms of about 10~m/s. When the transfer beam is turned on and the
additional repumper in the detection region is left off, most of
the atoms are pumped into the $F=1$ dark state and traverse the
detection region undetected. The weak remnant signal in
Fig.~\ref{darkstate} corresponds to a small fraction of atoms that
are still in the bright state $F=2$. The presence of those atoms
is due to stray repumper light throughout the chamber that
accidentally returns some atoms from the dark into the bright
state before they leave the detection region.

The dashed curve in Fig.~\ref{darkstate} shows the signal observed
when both the transfer beam and the additional repumper beam in
the detection region are used. In that case, the acceleration of
the atoms is suspended at the leading edge of the transfer beam,
and resumes at the leading edge of the repumper beam in the
detection region. The fluorescence associated with the renewed
acceleration is measured. The peak of this signal is delayed by
0.4~ms with respect to the signal measured without transfer beam.
Assuming that the transfer beam is located approximately 3~mm from
the MOT, from the observed delay of 0.4~ms we deduce an average
velocity of the dark beam of about 5~m/s. This is more than a
factor of two slower than any atomic beam based on the LVIS
design. The velocity of the dark atomic beam can be controlled by
the intensity of the push beam as well as the distance of the
transfer beam from the MOT center. We demonstrate this in the next
two sections.

\subsection{Effect of the Push Beam Intensity}

In the following, we discuss a series of fluorescence measurements
obtained for various push beam intensities. The rising edge of the
transfer beam is located 3~mm from the MOT center. The MOT beams
are always on, and the radiative force of the pulsed push beam
must overcome the damping force of the MOT to extract the atoms.
We find that for a given push beam intensity the push beam is most
effective if it is $\sigma$-polarized, with a helicity that is
opposite to that one would use if the pusher beam was a MOT trap
beam. Under this condition, the magnetic field of the MOT tunes
the atoms closer to resonance with the push beam as they are
pushed out from the MOT, resulting in a larger acceleration. We
have also found that for the push beam to overcome the damping
force of the MOT, the push beam power must exceed a threshold
power of about 0.1~mW. Below this power, the push beam merely
displaces the MOT, and there is no atomic beam.

\begin{figure} [pbt]
\centerline{\epsfig{figure=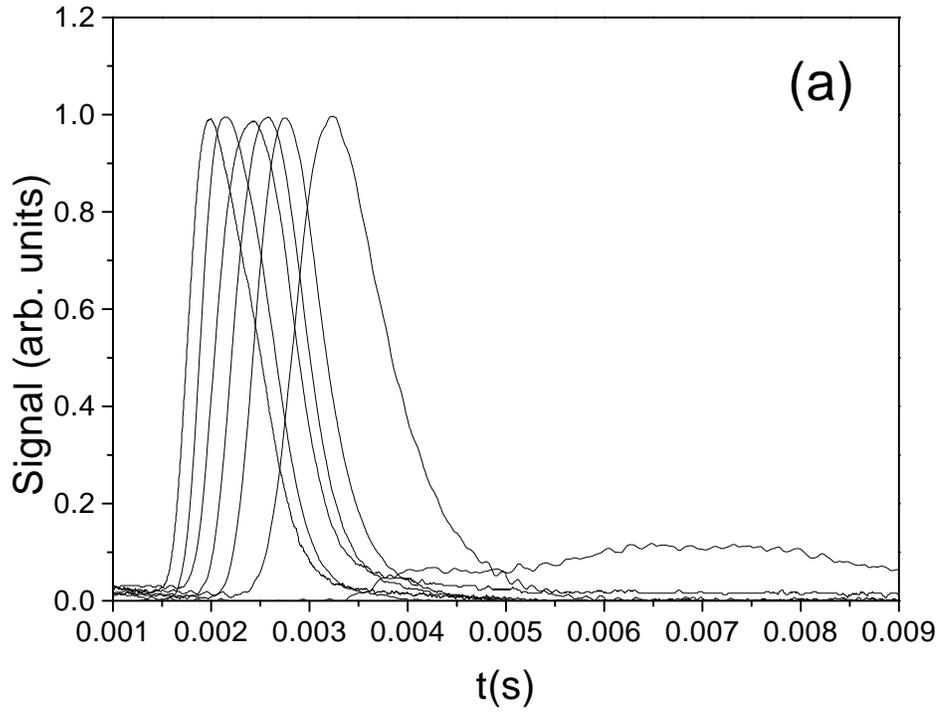,width=5.0in}}
\centerline{\epsfig{figure=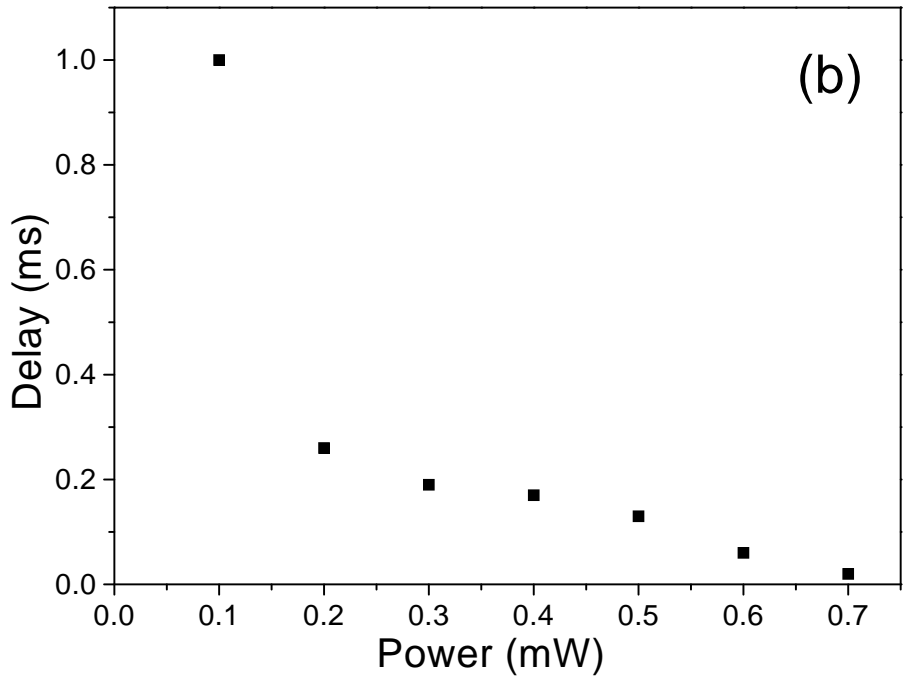,width=5.0in}}
\caption {(a) Fluorescence signals of pulsed atomic beams for push
beam powers ranging from 0.1~mW to 0.7~mW (right to left in steps
of 0.1~mW). The push beam diameter is 2~mm. Both the transfer
laser and the additional repumper laser in the detection region
are turned on. A minimum power of 0.1~mW is needed to extract
atoms. (b) Delay of most probable arrival time of atoms due to
effect of the transfer beam versus the power of the push beam.}
\label{intensityscan}
\end{figure}

Fig.~\ref{intensityscan}(a) shows the fluorescence signal for
different values of the push beam power when both the transfer
beam and the additional repumper laser in the detection region are
turned on. As the push beam power decreases, the fluorescence
signal peaks at later times, corresponding to a lower average beam
velocity. For each push beam power, we have, for comparison, also
measured the fluorescence signals without transfer laser (data not
shown). For the resultant pairs of curves, one pair being
analogous to the solid and the dashed curve of Fig.~1, we have
then determined the difference in the most probable arrival time
of the atoms. In Fig.~\ref{intensityscan}(b), the time differences
are plotted vs. push beam power. With decreasing power, the delay
time between the transfer-on and transfer-off signals increases
from 0.02~ms to 1.00~ms. This behavior reflects the fact that the
time of travel of the dark atomic beam from the transfer beam to
the detection region increases with decreasing velocity.

\subsection{Effect of the Transfer Beam Position}

\begin{figure} [hbt]
\centerline{\epsfig{figure=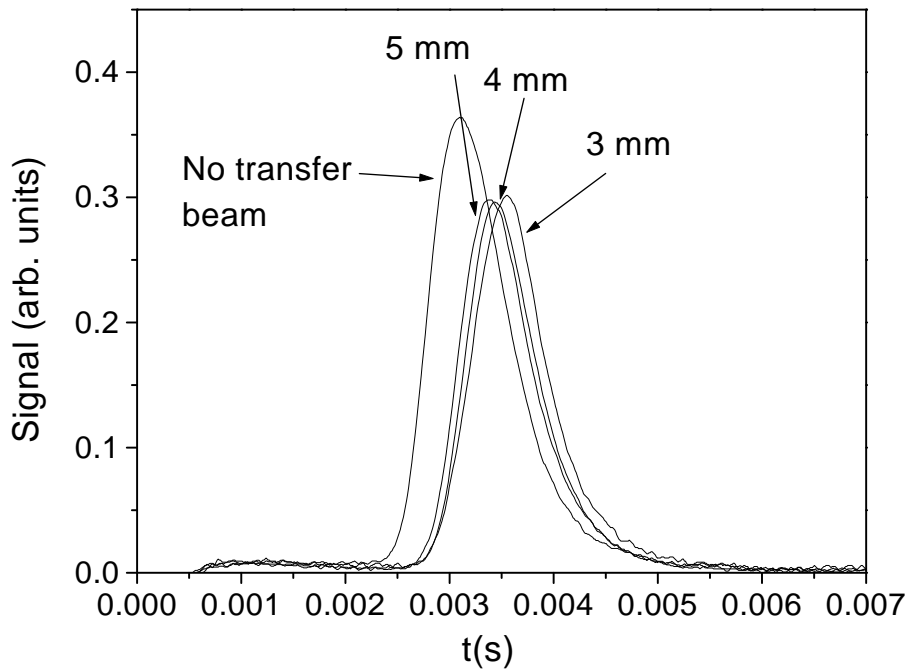,width=5.0in}} \caption
{Fluorescence signal measured for three indicated distances of the
transfer beam from the MOT center. The peaks of the fluorescence
curves arrive later as the transfer beam is moved closer to the
MOT center.} \label{posscan}
\end{figure}

The velocity of the dark atomic beam can also be controlled by
varying the position of the transfer beam relative to the MOT
center. Moving the transfer laser closer to the MOT leaves the
atoms less time to interact with the push beam before they are
pumped into the dark state. This  results in a slower dark atomic
beam. Fig.~\ref{posscan} shows the fluorescence signal, with both
the transfer beam and the additional repumper beam on, for three
distances of the transfer beam from the MOT. As explained in
Sec.~\ref{esetup} and in Fig.~1, the cutoff of the MOT repumper
beam and the transfer beam need to be varied such that there
always remains a small gap between both beams. As expected, the
delay time caused by the transfer beam increases when the transfer
beam is moved towards the MOT. There is a tradeoff between
attaining a large atom flux and a low atomic-beam velocity.  While
moving the transfer laser beam closer to the MOT results in a
slower atomic beam (which may be desired), it also entails the
necessity to move the MOT repumper cutoff closer to the MOT
center, thereby decreasing the MOT loading efficiency and the
atomic-beam flux.

\section{Numerical Simulation}

We have numerically simulated the dynamics of the MOT with the
push and transfer beams using a classical description of the
center-of-mass motion and a quantum-mechanical description of the
internal dynamics of the atoms. We have used the analysis given in
\cite{pushLVIS1} as a starting point for our model. For a
two-level system, the radiation pressure force acting on the atom
due to a single laser beam is the product of the photon momentum
multiplied by the photon scattering rate $\Gamma_{\rm scatter}$,
and is given by

\begin{eqnarray} {\bf{F}} = \hbar {\bf{k}} \Gamma_{\rm scatter}
 = \hbar {\bf{k}} \frac{\Gamma}{2} \frac{s}{s+1} \, ,
\label{semiforce}
\end{eqnarray}

where $\Gamma$ is the decay rate of the excited state and $s$ is
the saturation parameter

\begin{eqnarray} s=\frac{I}{I_{\rm sat}}\frac{{\Gamma}^2}{{\Gamma}^2 +
4(\delta-{\bf{k}}\cdot {\bf{v}} - \Delta E_{\rm Zeeman}/\hbar)^2}
\,.
\end{eqnarray}

The saturation parameter takes into account the Doppler shift due
to the (classical) atomic velocity ${\bf{v}}$  as well as the
relative Zeeman shift $\Delta E_{\rm Zeeman}$ between the involved
states arising from the magnetic field of the MOT coils and the
atom guide wires.  Further,  $\delta$ is the field-free laser
detuning, $I$ the beam intensity, and $I_{\rm sat} =
1.6$~mW/cm$^2$ the saturation intensity.

The $F=2 \rightarrow F'=3$ MOT transition of $^{87}$Rb involves
multiple magnetic subcomponents with respective Zeeman shifts and
Clebsch-Gordan coefficients. Further, to be able to calculate the
average radiation pressure force, the probability distribution of
the atom over the magnetic sublevels of the $F=2$ ground state
needs to be determined. To make the model numerically more
tractable, we use an $F=0 \rightarrow F'=1$ model transition with
wavelength and saturation intensity equal to those of the Rb D2
line, as explained in \cite{pushLVIS1}. Also, due to the low
density of atoms in MOTs designed to produce cold atomic beams, we
can ignore collisions between atoms and simulate the dynamics of
each particle independently.

Using these assumptions, the radiation-pressure force on an atom
due to the MOT beams and the pusher beam becomes

\begin{eqnarray}
{\bf{F}}=\sum_{i=1,7}\sum_{m=-1,1}\hbar{\bf{k}}_{\rm i}
\frac{\Gamma}{2} \frac{s_{\rm i,m}} {1+\sum\limits_{i',m'} s_{\rm
i',m'}} \label{forceeqn}
\end{eqnarray}

where

\begin{eqnarray}
s_{\rm i,m}=\frac{I_{\rm i,m}}{I_{\rm sat}} \frac{\Gamma^2}
{\Gamma^2+4(\delta_{\rm i}-{\bf{k}}_{\rm i}
\cdot{\bf{v}}+m\mu_{\rm B} / \hbar)^2}
\end{eqnarray}

is the saturation parameter of the individual beam components. The
$i$ and $m$ are indices for the beam number (six MOT beams plus
one push beam) and the laser polarization, respectively. Each
laser beam $i$ is decomposed into its $\sigma_-$, $\pi$ and
$\sigma_+$ intensity components $I_{\rm i,-1}, I_{\rm i,0}$, and
$I_{\rm i,-1}$ in a coordinate frame aligned with the
magnetic-field direction at the (time-dependent) location of the
atom. The denominator term $1+\sum\limits_{i',m'} s_{\rm i',m'}$
in Eq.~\ref{forceeqn} is introduced in order to obtain a
reasonable saturation behavior \cite{pushLVIS1} (the magnitude of
the force on an atom must not exceed $\hbar k \Gamma/2$).

\begin{figure} [hbt]
\centerline{\epsfig{figure=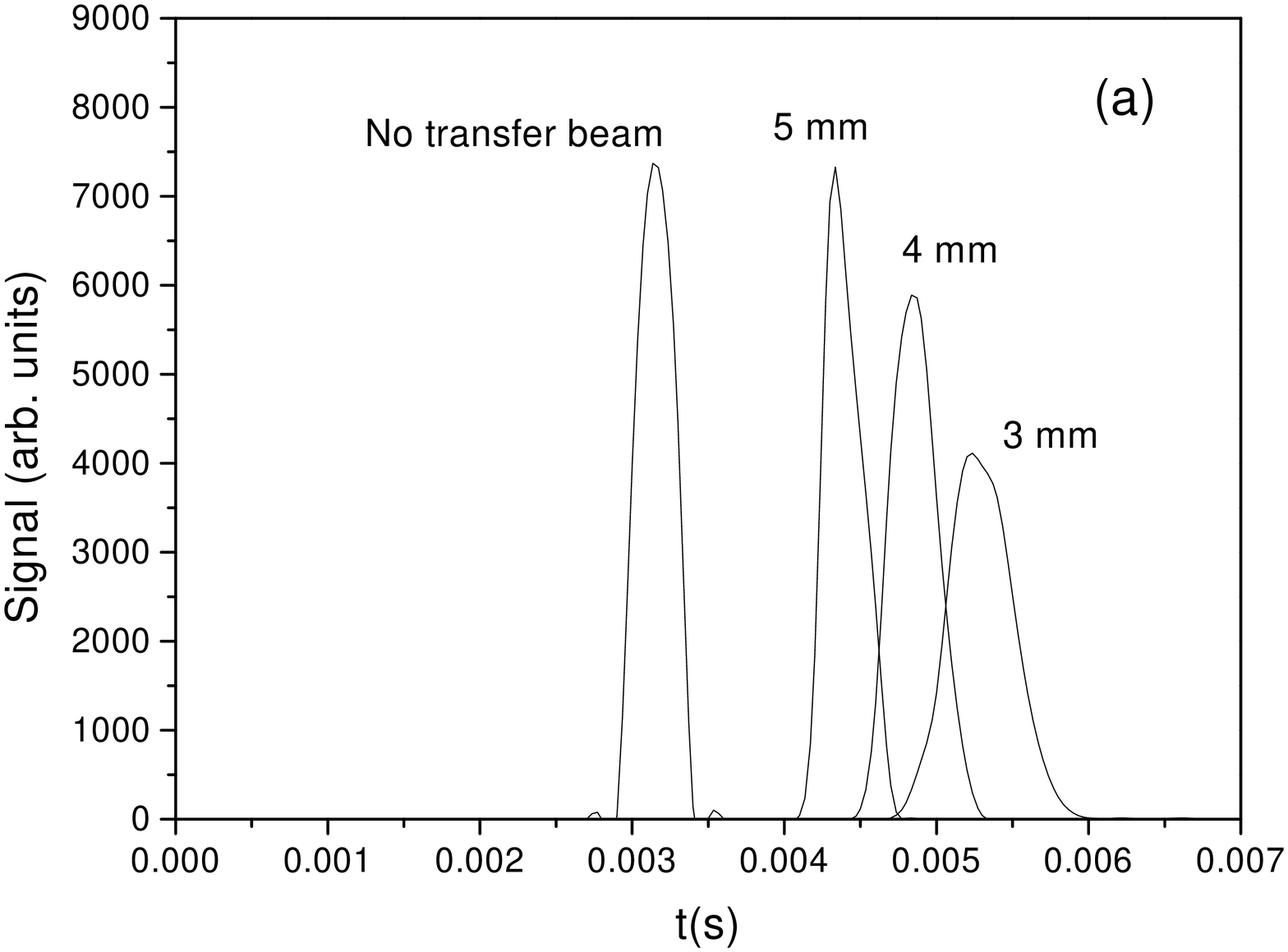,width=4.5in}}
\centerline{\epsfig{figure=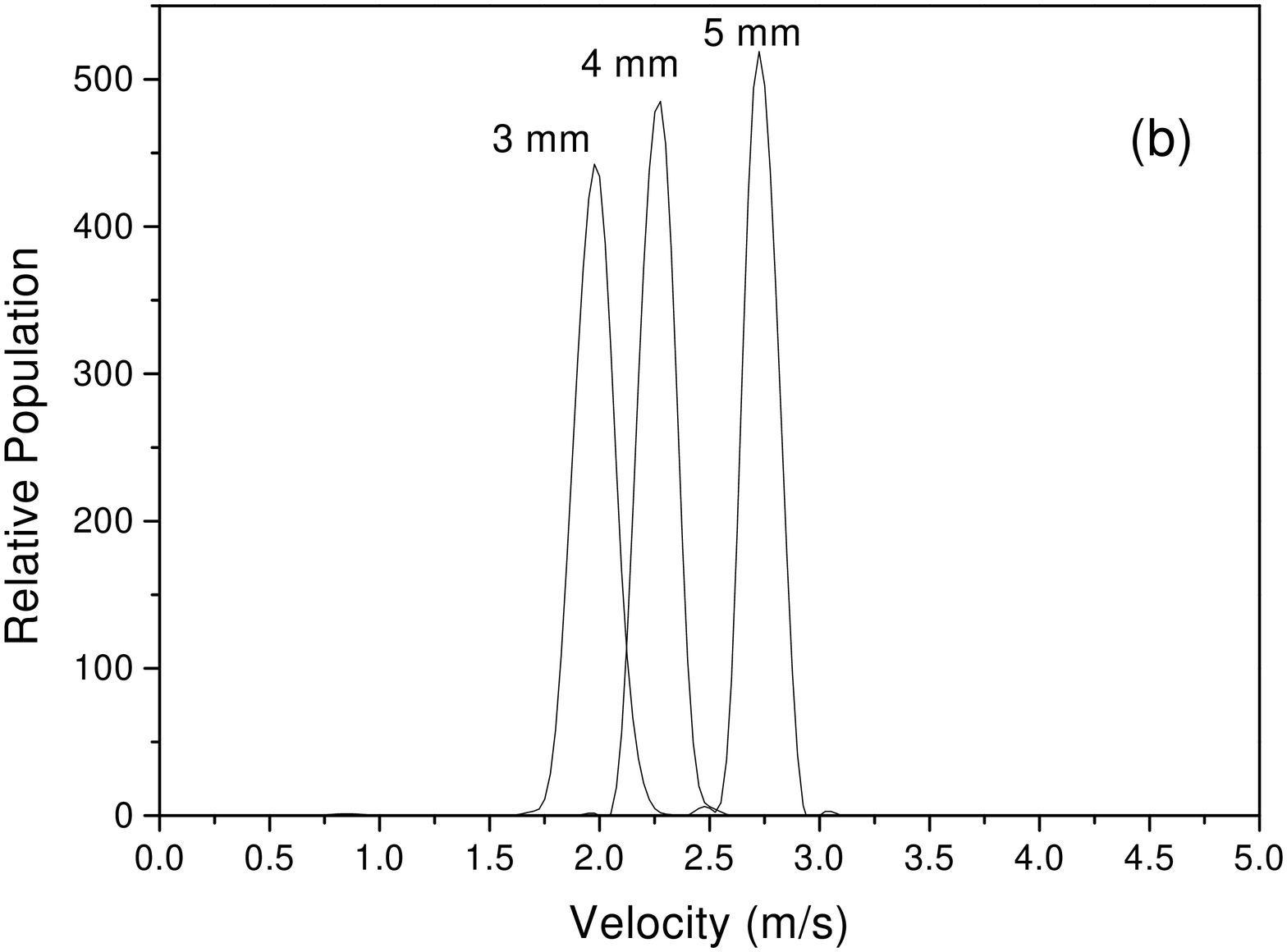,width=4.5in}} \caption {(a)
Simulated fluorescence signals corresponding to the experimental
parameters in Fig.~4 for transfer beams at 3, 4 and 5~mm from the
MOT center. Push beam intensity is 8~mW/cm$^2$ and single-beam MOT
intensity is 5~mW/cm$^2$. (b) Corresponding distributions of the
velocity of atoms in the dark atomic beam.}
\label{sim2}
\end{figure}

In the following, we explain the features we have added to the
model outlined in \cite{pushLVIS1}. Spontaneous emission causes a
diffusive motion in momentum space. In our model, we integrate, as
the atom propagates, the total number of photons $N$ that have
been scattered. At suitable time intervals, a velocity kick of
magnitude $\sqrt{N}$ times the single-photon recoil velocity and
with a random direction is added to the velocity of the atom, and
$N$ is re-set to zero. We have verified that this procedure
produces the expected steady-state temperature and cloud size of
the MOT (with the push beam turned off). Further, in addition to
the ground and excited states, we introduce a virtual dark state
that the atom can be transferred to by a transfer beam. This state
does not interact with the MOT or push beams. The role of the
repumper is simulated by a transfer of atoms back to the $F=0$
state of the simulation. We ignore the momentum changes associated
with the repumper and transfer beams because they amount to just a
few photon recoils.

Fig.~\ref{sim2}(a) shows the simulated fluorescence signals
corresponding to the conditions for the data shown in
Fig.~\ref{posscan}. The calculations reproduce the qualitative
features of the effect of the push beam position. The
corresponding velocity distributions of the atoms are shown in
Fig.~\ref{sim2}(b). The simulations indicate a much lower beam
velocity than the experimental data. This quantitative discrepancy
can be attributed in part to imperfect beam geometries in the
experiment, and, mostly,  to  repumper stray light scattered by
the uncoated surfaces of our quartz vacuum chamber. The repumper
stray light can re-pump atoms that have been transferred into the
dark state back into the bright state and thereby make them
susceptible to unwanted acceleration.

\begin{figure} [hbt]
\centerline{\epsfig{figure=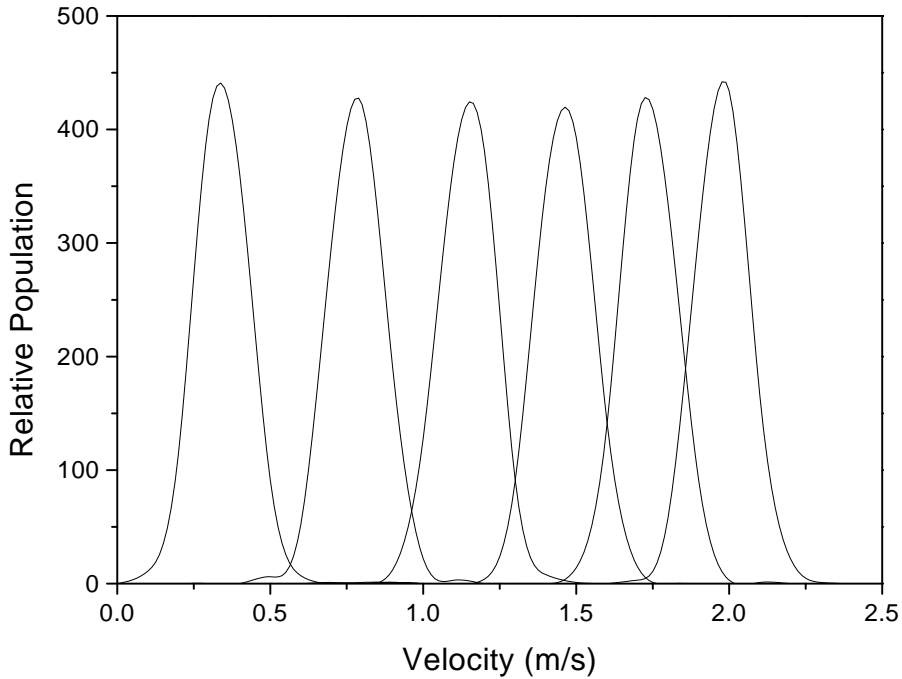,width=5.0in}} \caption
{Simulated velocity distributions for push beam intensities
ranging from 3 to 8~mW/cm$^2$ (in steps of 1~mW/cm$^2$ from left
to right) and a transfer beam located 3~mm from the MOT center.
The single-beam MOT intensity is 5~mW/cm$^2$.} \label{sim1}
\end{figure}

The atomic beam velocity is sensitive to all laser beam
parameters. In our simulation, suitable parameters lead to dark
atomic beams with velocities as slow as 0.3~m/s, which is about a
factor 30 smaller than the method of the LVIS. A few simulated
velocity distributions are shown in Fig.~\ref{sim1}. There, the
RMS velocity spread is of order 0.2~m/s in all cases. The slowest
velocites are obtained using push beam intensity of 3~mW/cm$^2$
and a transfer beam separation from the MOT of 3~mm. While
experimental limitations have prevented us from a quantitative
reproduction of these theoretical results, we have been able to
demonstrate the basic functionality of the technique.

\section{Conclusion}

We have experimentally demonstrated  a simple technique to produce
atomic beams with velocities and velocity spreads that are much
smaller than that obtained from an LVIS. By pumping the atoms into
a dark state, we have produced a slow, dark atomic beam that can
be guided in a magnetic atom guide. Since the guided, dark atomic
beam is not susceptible to stray light from the nearby MOT atom
source, it can be operated continuously and concurrently with the
MOT. Our numerical simulations show that the technique can produce
dark atomic beams with velocities less than 0.2~m/s and velocity
spreads comparable to that of a (doppler-limited) moving optical
molasses. Our small vacuum chamber with uncoated windows has
resulted in significant stray repumper light in the guide, which
prevented us from observing the very slow velocities predicted by
our simulations. This problem can, however, easily be alleviated
using a larger vacuum chamber with fewer reflecting surfaces. We
believe that dark, guided atomic beams are an ideal atom source
for the realization of a continuous BEC in a magnetic atom guide.

\end{document}